\documentclass[usenatbib]{mn2e}

\usepackage{bm}
\usepackage{latexsym}
\usepackage{eufrak}
\usepackage{epsfig}
\usepackage{latexsym}
\usepackage{amsmath}
\usepackage{epsfig}
\usepackage{ifpdf}
\usepackage{graphicx}
\usepackage{dcolumn}
\usepackage{psfrag} 
\usepackage{booktabs}

\def\beq{\begin{equation}}
\def\eeq{\end{equation}}
\def\bea{\begin{eqnarray}}
\def\eea{\end{eqnarray}}

\usepackage{natbib}
\title[Bayesian study of the primordial power spectrum]{A Bayesian study of the primordial power spectrum 
from a novel closed universe model}

\author[V\'azquez et al.]
{J.~Alberto~V\'azquez$^{1,2}$\thanks{E-mail:jv292@cam.ac.uk}, A.N.~Lasenby$^{1,2}$, 
M.~Bridges$^{1,2}$, M.P.~Hobson$^{1}$\\
$^{1}$Astrophysics Group, Cavendish Laboratory, JJ Thomson Avenue, Cambridge CB3 0HE, UK.\\
$^{2}$Kavli Institute for Cosmology, Madingley Road, Cambridge CB3 0HA, UK.}

\begin{document}

\date{Accepted --. Received -- ; in original form 11 December 2011}
\pagerange{\pageref{firstpage}--\pageref{lastpage}} \pubyear{2011}
\maketitle

\label{firstpage}

\begin{abstract}
We constrain the shape of the primordial power spectrum using recent
measurements of the cosmic microwave background (CMB) from the
Wilkinson Microwave Anisotropy Probe (WMAP) 7-year data and other
high-resolution CMB experiments. We also include observations of the
matter power spectrum from the luminous red galaxy (LRG) subset DR7 of
the Sloan Digital Sky Survey (SDSS).  We consider two different models
of the primordial power spectrum.  The first is the standard nearly
scale-invariant spectrum in the form of a generalised power-law
parameterised in terms of the spectral amplitude $A_{\rm s}$, the
spectral index $n_{\rm s}$ and (possibly) the running parameter $n_{\rm
  run}$. The second spectrum is 
derived from the Lasenby and Doran (LD) model. The LD model is based on
the restriction of the total conformal time available in a closed
Universe and the predicted primordial power spectrum depends upon 
just two parameters.  An important feature of the LD spectrum is that it
naturally incorporates an exponential fall-off on large scales, which
might provide a possible explanation for the lower-than-expected power
observed at low multipoles in the CMB.  In addition to parameter
estimation, we compare both models using Bayesian model selection.  We
find there is a significant preference for the LD model over a simple
power-law spectrum for a CMB-only dataset, and over models with an equal 
number of parameters for all the datasets considered.
\end{abstract}

\begin{keywords}
cosmological parameters -- cosmology: observations -- cosmology: theory -- 
cosmic background radiation -- large-scale structure 
\end{keywords}

\section{Introduction}

Cosmological inflation not only explains the homogeneity of the universe on
large scales, but also provides a theory for explaining the observed
level of anisotropy \citep{Guth81,Albrecht82, Mukhanov82}.
Inflationary models generically predict Gaussian, adiabatic and nearly
scale-invariant primordial fluctuations.
To determine the shape of the primordial power spectrum from
cosmological observations, it is usual to assume a parameterised form
for it.  The simplest assumption is that the initial spectrum has the
form of a simple power-law, parameterised in terms of the spectral
amplitude $A_{\rm s}$ and the spectral index $n_{\rm s}$. Recent
analyses have shown, however, that the spectral index may deviate
from a constant value (close to unity) and so
consideration of models that provide some \emph{running} of the index
(defined by $n_{\rm run} \equiv d n_{\rm s} / d \ln k$) is warranted.
Furthermore, models that predict a decrement in CMB power at low
multipoles seem to be preferred by observations \citep{Efstathiou03,
  Contaldi03}, so that, an extra variable to account for
a cut-off scale might be considered.  It would be feasible to continue
 adding parameters in this fashion until some arbitrary accuracy of
model fit is achieved, but this procedure would fail to account for
Occam's razor: a simpler model should be preferred, unless the data
require a more sophisticated one. Using the Bayesian evidence to
select between models is one way to include this consideration.

There have been several recent studies regarding the shape of the
primordial spectrum, some based on physical models, some using
observational data to constrain an \textit{a priori} parameterisation,
and others attempting a direct reconstruction from data
\citep{Barriga01,Bridle03, Hannestad04, Bridges05,
  Bridges06,Bridges08, Verde08, Peiris10, Hlozek11, Guo11}.  
 A primary goal of this work is to fit the
\emph{predicted} form of the primordial spectrum from a physically
motivated model described by \citet{Lasenby05} (henceforth LD). 
We translate observational data into
constraints on the generalised power-law and LD forms for the
primordial power spectrum (and the standard cosmological parameters),
and decide which model provides the best fit to observational data
using the Bayesian evidence.

The paper is organised as follows: in Section~2 we present the two
different models for the power spectrum and in Section~3 we describe
basic parameter estimation and model selection. We list the datasets
and the cosmological parameters considered in Section~4 and present
the resulting parameter constraints in Section~5. We compute
Bayesian evidences in Section~6 to decide which model provides the
best description for current observational data and we validate our analysis by
applying it to a CMB simulated data. Our conclusions are presented in Section~7.

\section{Primordial power spectrum}

The {\it correlation function} $\xi$ of density fluctuations $\delta\equiv\delta\rho/\rho$
at two separated points $\bmath{x}$ and $\bmath{x}+\bmath{r}$ is defined as 
\begin{equation}\label{eq:two-point}
   \xi(r) \equiv  \langle \delta(\bmath{x}) \delta(\bmath{x}+\bmath{r}) \rangle.
\end{equation}
Because the assumption of
homogeneity and isotropy, $\xi$ is a function only of $r \equiv
|\bmath{r}|$.
The {\it power spectrum} $\mathcal{P}(k)$ describes 
the amplitude of fluctuations on different length scales
and it is related with the inverse Fourier transform of the 
correlation function $\xi$ by:
\begin{equation}\label{eq:Fourier}
  \mathcal{P}(k) \equiv  \langle |\delta_k |^2  \rangle. 
\end{equation}
During the inflationary period, fluctuations in the inflaton field $\delta \phi_k$ result 
in curvature perturbations $\mathcal{R}(k)$ given by 
\begin{equation}\label{eq:curv}
\mathcal{R}(k) = -\left[ \frac{H}{\dot \phi} \delta \phi_k\right]_{k= RH},
\end{equation}
where the quantities are evaluated at the horizon exit epoch
$k=RH$. Here $R$ is the scale factor of the universe and $H\equiv
\dot{R}/{R}$ is the Hubble parameter.
In this paper, we follow a slow-roll approximation \citep[e.g.][]{Liddleboo}. 
Hence the power spectrum of the inflaton fluctuations is constant in time and
equal to
\begin{equation}\label{eq:phi_spectrum}
  \mathcal{P}_{\delta \phi}(k) = \left(\frac{H}{2\pi}\right)^2_{k=RH}.
\end{equation}
Thus, the 
{\it primordial curvature spectrum} $\mathcal{P_R}(k)$ computed from 
(\ref{eq:Fourier}) - (\ref{eq:phi_spectrum}) is
\begin{equation}\label{eq:spectrum}
  \mathcal{P_R}(k) =\left[ \left(\frac{H}{\dot \phi}\right)^2 \left(\frac{H}{2\pi}\right)^2 \right]_{k=RH}.
\end{equation}

\subsection{Power-law parameterisations}

\noindent
Cosmological slow-roll inflation predicts the spectrum of curvature perturbations
to be close to scale-invariant.
Based on this, the spectrum is
commonly assumed to have the form
\begin{equation}
  \mathcal{P_R}(k)=A_{\rm s} \left( \frac{k}{k_0}  \right)^{n_{\rm s}-1},
\end{equation} 
where the {\it spectral index} $n_{\rm s}$ is expected to be close to
unity; $k_0$ is the pivot scale (set to $k_0= 0.05$ $\rm{Mpc}^{-1}$
throughout).
A spectrum where the typical amplitude of perturbations is identical
on all length scales is known as Harrison-Zel'dovich spectrum ($n_{\rm
  s}=1$).This particular parameterisation involves only one free
parameter, the {\it spectral amplitude} $\mathcal{P}(k)= A_{\rm s}$.

A further extension is possible by allowing the spectral index to vary
as a function of scale, such that $n_{\rm s}(k)$.  This can be achieved by
including a second order term in the expansion of the power spectrum
\begin{equation}
 \mathcal{P_R}(k) = A_{\rm s} \left( \frac{k}{k_0} \right)^{n_{\rm s}-1+(1/2)\ln (k/k_0)(dn/d\ln k)},
\end{equation}
where $dn/d \ln k$ is termed the {\it running parameter} $n_{\rm run}$ and
we would expect $n_{\rm run} \approx 0$ for standard inflationary models.

In what follows we will consider three Power-Law parameterisations. In
the first model (PL 1), we will assume a simple power-law
spectrum (without running) and restrict the universe to be spatially
flat. In PL model 2, we allow the spatial curvature of the universe to
be a free parameter. In PL model 3, we allow for a running spectrum,
but again restrict the universe to be spatially-flat. In this way, power-law 
models 2 and 3 have the same number of parameters.

\subsection{The LD model} 

Assuming the cosmological constant is the origin of dark energy,
\citet{Lasenby05} provided a construction for embedding
closed-universe models in a de Sitter background.  As a consequence of
this novel approach, a boundary condition on the total available
conformal time emerges. Defining the total conformal time $\eta$ as 
\begin{equation}\label{eq:eta}
  \eta \equiv \int^\infty_0 \frac{dt}{R(t)},
\end{equation}
the LD model requires $\eta=\pi/2$.  For more details about the choice
of the boundary condition, including how it can be reinterpreted as an
eigenvalue condition on the solution of a differential equation, see
\citet{Lasenby03, Lasenby04, Lasenby05}. 
In order to understand some consequences of the new boundary condition
we split the history of the Universe in two main contributions to the
total conformal time: matter (radiation and dust) and inflationary eras.
Hence, we want to compute the conformal time $\eta_{\rm M}$ elapsed during
the matter era and add it to that elapsed in the inflationary era
$\eta_{\rm I}$, such that the boundary condition is satisfied:
 \begin{equation}\label{eq:constraint}  
   \eta_{\rm I}+\eta_{\rm M}=\frac{\pi}{2}.
 \end{equation}  
It is found that this constraint leads to a `see-saw' mechanism linking the
parameters describing the current state of the universe with the initial conditions \citep{Lasenby04}.

 \subsubsection{Matter era}
 
The general description of the large scale Universe is based on the
Robertson-Walker space-time with dynamics governed by the Einstein
equations. The resulting Friedmann equations can be written as (with
$c=1$)
\begin{eqnarray}\label{eq:fried}
   \frac{\dot R^2 + k}{R^2}-\frac{\Lambda}{3} &=& \frac{8\pi G}{3} \, \rho,  \\
   2\frac{\ddot{R}}{R} + \frac{\dot R^2 + k}{R^2}- \Lambda &=& -8\pi G P. \nonumber 
\end{eqnarray}
Here $k = 0$, $\pm 1$ defines the geometry of the universe, $\Lambda$ is
the cosmological constant, and the relationship between density $\rho$
and pressure $P$ is encoded in the equation of state $P=\gamma \rho$.
The behaviour of the homogeneous universe is governed by the parameters representing 
its matter-energy content, namely
\begin{equation}
   \Omega_{M}= \frac{8\pi G \rho}{3H^2} ,\qquad \qquad    \Omega_{\Lambda}= \frac{\Lambda}{3H^2}, 
\end{equation}
and its expansion history defined by the Hubble parameter $H$.
Moreover, if we assume the matter density is made up of decoupled dust
and radiation, the equations  governing $\Omega_M$ and $\Omega_{\Lambda}$ can
be solved exactly and its solution is controlled by two arbitrary
constants $\alpha$ and $\beta$ given by
\begin{eqnarray}\label{eq:alpha}
    \alpha &=& \frac{\Omega_{m_o}^2 \Omega_{\Lambda_o}}{(\Omega_{m_o}+ \Omega_{r_o} + \Omega_{\Lambda_o} -1)^3}, \\
    \alpha \beta  &=& \frac{\Omega_{r_o} \Omega_{\Lambda_o}}{(\Omega_{m_o}+ \Omega_{r_o} + \Omega_{\Lambda_o} -1)^2}, \nonumber
\end{eqnarray}   
where subscript `$o$' denotes quantities  evaluated  at present time.
The total conformal time for this type of universe can be written in terms of the dimensionless parameters  
$\alpha$ and $\beta$ as   \citep[see][]{Lasenby05} 
\begin{equation}\label{eq:eta_matt}
\eta_M =     \int^{\infty}_0 \frac{dx}{(\beta x^4 + x^3 - x^2 +\alpha)^{1/2}}.  
\end{equation}  
%

\subsubsection{Inflationary era}

 The computation of the conformal time in the inflationary epoch is a
 more elaborate process.  Let us consider a basic inflationary model
 where the particle responsible for this process is simply a real,
 time-dependent, homogeneous, free, massive scalar field $\phi$, 
 described by the equations
\begin{eqnarray} \nonumber
  \dot H + H^2 - \frac{\Lambda}{3} + \frac{4\pi G}{3}(2\dot  \phi^2 - m^2\phi^2 ) &=&  0,\\
  \ddot{\phi} + 3H\dot \phi + m^2 \phi &=&  0. \label{eq:inf}
\end{eqnarray} 
For closed universe models, the scale factor is given explicitly by
\begin{equation}\label{eq:inf2}
   \frac{1}{R^2} = \frac{4\pi G}{3}(\dot \phi^2 + m^2 \phi^2) - H^2 + \frac{\Lambda}{3}.
\end{equation} 
 In order to compute the conformal time $\eta_{I}$, it is necessary
 to seek out suitable conditions before the onset of inflation and
 then solve the dynamics for the scalar field encoded in equations
 (\ref{eq:inf}). To do this, \citet{Lasenby05}
 developed formal series expansions out of the initial singularity,
 $t=0$, in terms of dimensionless
 variables $u=t/t_{\rm p}$ and $\mu=m/m_{\rm p}$, where the subscript `p'
 denotes a Planck units variable. The series are given by
\begin{eqnarray}
\phi(u)&=&\frac{1}{l_{\rm p}}\sum_{n=0}^{\infty}\phi_n(u)\ln^n(u), \nonumber\\
H(u)&=& \frac{1}{t_{\rm p}}\sum_{n=0}^{\infty}H_n(u)\ln^n(u), 
\end{eqnarray}

\noindent
with
\begin{eqnarray}\label{eq:expansion}
 \phi_0&=&b_0+b_4 u^{4/3}-\frac{118\sqrt{3\pi}b_4^2}{99}u^{8/3}
      -\frac{u^2}{1296\pi}(11\sqrt{3\pi}\mu ^2 \nonumber\\ 
       & & - 54\sqrt{3\pi}\Lambda -216\sqrt{3}\pi^{3/2} \mu^2 b_0^2 + 36\pi \mu^2 b_0 ), \nonumber \\  
  \phi_1&=&-\sqrt{\frac{1}{12\pi }}-\frac{\mu ^2}{216\pi }\left(-\sqrt{3\pi }+36\pi b_0\right)u^2, \nonumber \\
  H_0&=&\frac{1}{3u}+\frac{32\sqrt{3\pi}}{27}b_4 u^{1/3} +(\frac{2\mu ^2}{81}+\frac{\Lambda }{3}+\frac{4\pi}{3}\mu ^2 b_0^2 
  \nonumber \\
      && +  \frac{4\sqrt{3\pi}}{27}\mu ^2 b_0 )u - \frac{6656\pi b_4^2}{891}u^{5/3}, \nonumber \\           
  H_1&=&-u \frac{dH_0}{du} - u H_0^2+\frac{u\Lambda }{3}-\frac{8\pi\ u}{3}\left( \frac{d\phi_0}{du} \right)^2 \nonumber\\
 	&& -\frac{16\pi \phi_1}{3} \frac{d	\phi_0}{du} -\frac{8\pi \phi_1^2}{3u}+\frac{4\pi \mu ^2 u \phi_0^2}{3}.  
\end{eqnarray}
   
 We observe that two new free parameters $b_0$ and $b_4$, appear in
 the series expansions in (\ref{eq:expansion}).  Together with the mass of the scalar  field $\mu$,
 they control the magnitude of the field and how long the inflationary
 period lasts.  In order to decide on the priors we shall employ in
 our subsequent Bayesian analysis, it is worth pointing out some
 features related with these new parameters.
\begin{itemize}    
  \item The amplitude of the perturbations is determined by the scalar field mass $\mu$. To match the observed level of CMB anisotropies, 
  we shall need to set it to be about 
  \begin{equation}
\mu \sim 10^{-6}.
\end{equation}  
\item The number of e-foldings $N$ is primarily determined by $b_0$ and may be approximated as 
    \begin{equation}\label{eq:Nb0}
      N  \approx 2\pi b_0^2. 
    \end{equation}
Hence, to obtain realistic models we need $b_0$ to be of order $\sim$ a few.
\item The conformal time is estimated by 
    \begin{equation}\label{eq:eta_I}
      \eta_{I} \approx 0.92 \left(\frac{|b_4|}{\mu^{4/3}} \right)^{1/2}\left(\frac{1}{b_0^2}\right).
    \end{equation}
Employing the constraint (\ref{eq:constraint}), $|b_4|\mu^{-4/3}$
should thus be around unity.
   
\item The parameter $b_4$ controls the initial curvature, as can be
  seen from (\ref{eq:inf2}):
     \begin{equation}\label{eq:ini_cur}
        \frac{R}{l_{\rm p}} \approx \left(\frac{2187}{12544\pi}\right)^{1/4} \frac{u^{1/3}}{\sqrt{-b_4}}.
     \end{equation}
Therefore $b_4$ must be negative. Making use of the rest of the
parameters and (\ref{eq:eta_I}), $|b_4|$ should be around
$10^{-9}$.
\end{itemize}
The restriction on the values for the model parameters together with
the boundary condition, severely limits the class of models allowed to
reproduce current cosmological observations. 

\begin{figure}
   \centerline{ \epsfxsize=210pt \epsfbox{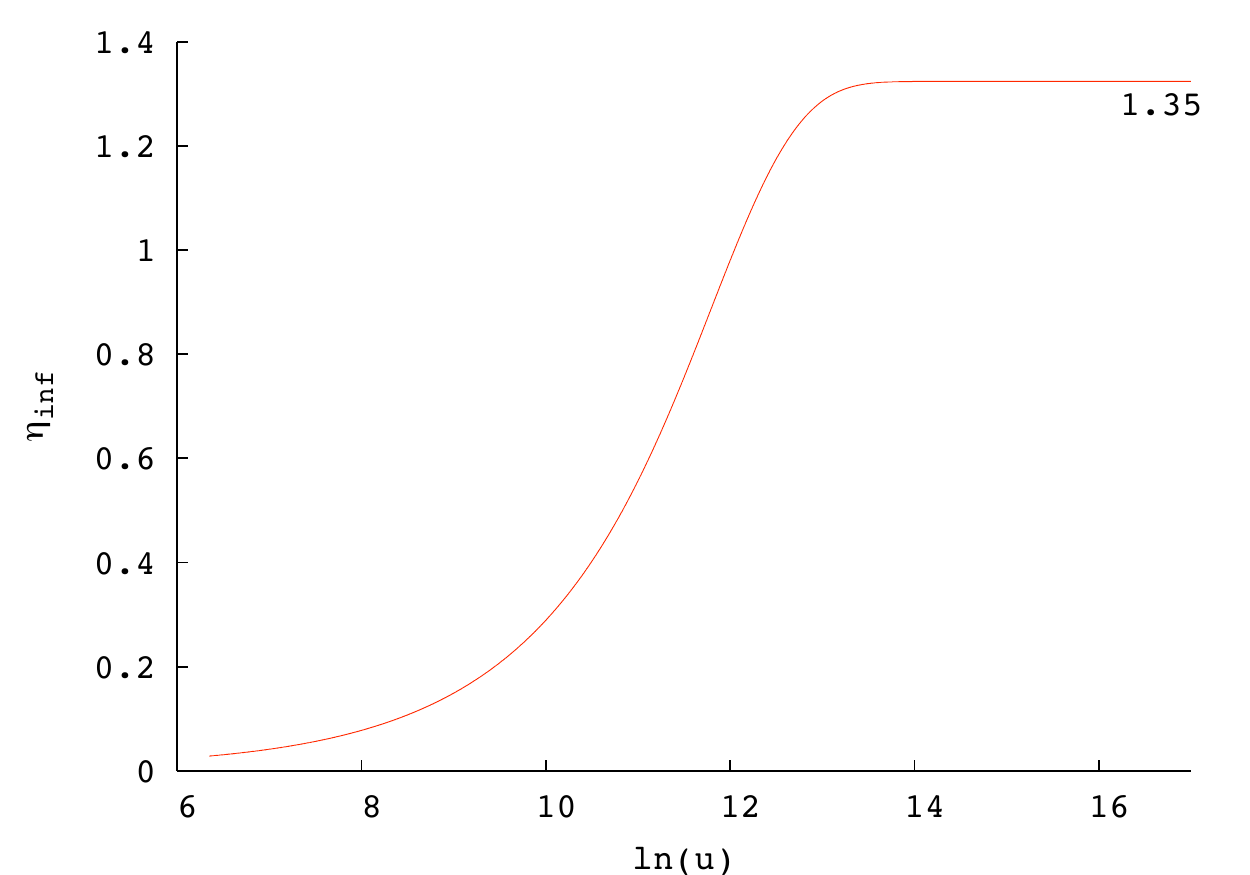} }
	\caption{Evolution of the conformal time $\eta_I$ as a function of $\ln(u)$. We observe that $\eta_I$ saturates at a value of around
	 1.35 by the end of inflation. The parameters used in this model are $b_0=2.47$, $b_4= -17.7\times 10^{-9}$
	  and $\mu=1.68\times 10^{-6}$.}
\label{fig:eta}
\end{figure}
  
Once we have found the initial conditions (\ref{eq:expansion}), it is
straightforward to solve numerically the dynamics of the scalar field
$\phi$ and the expansion history $H$ to determine the evolution of the
universe.  As an example, let us consider the best-fit values for the
cosmological parameters given by WMAP+BAO+$H_0$ \citep{Komatsu10a}, in
which case the conformal time elapsed during the matter epoch is
$\eta_M=0.22$.  In order to satisfy the boundary condition we should
choose appropriate values $\{b_0,b_4,\mu\}$ such that by the end of
the inflationary period the achieve conformal time is $\eta_{I}\approx 1.35$, as
shown on Fig \ref{fig:eta}. The expected shape of the primordial
power spectrum $\mathcal{P}(k)$ for this model is directly computed
from (\ref{eq:spectrum}).

\section{Bayesian analysis}

\subsection{Parameter Estimation}

A Bayesian analysis provides a consistent approach to estimate the
set of parameters $\mathbf{\Theta}$ within a model $M$, which
best describes the data $\bmath{D}$. The method is based on the
assignment of probabilities to all quantities of interest, and then
the manipulation of those probabilities given a series of rules, of
which \textit{Bayes' theorem} plays the principal role,
\citep{Bayes63}.  Bayes' theorem states that
\begin{equation}\label{eq:bayes}
  P( \mathbf{\Theta}|\bmath{D},M)= \frac{P(\bmath{D}|\mathbf{\Theta},M) \,\, P(\mathbf{\Theta}|M)}{P(\bmath{D}|M)},
\end{equation}
where the {\it prior} probability $P(\mathbf{\Theta}|M) \equiv
\pi$ represents our knowledge of $\mathbf{\Theta}$ before considering
the data. This probability is modified through the {\it likelihood}
$P(\bmath{D}|\mathbf{\Theta},M) \equiv \mathcal{L}$ when the
experimental data $\bmath{D}$ are considered. The final objective for
Bayesian inference is to obtain the posterior probability
$P(\mathbf{\Theta}|\bmath{D},M)$ which represents the state of
knowledge once we have taken the new information into account. The
normalisation constant is the marginal likelihood or {\it Bayesian
 evidence} $P(\bmath{D}|M) \equiv \mathcal{Z}$. 
 Since this quantity is independent of the
parameters $\mathbf{\Theta}$, it is commonly ignored in parameter
estimation but takes the central role for model selection \citep{Hobson02, Liddle04}.

\subsection{Model selection}

\noindent
A complex model that explains the data slightly better than a simple
one should be penalised for the introduction of extra parameters,
because the additional parameters bring with them a lack of
predictability. Also, if a model is too simple, it might not fit the
data equally well, then it can be discarded \citep[e.g.][]{Liddle06,  Liddle07, Trotta08}.
Following this line, many attempts have been performed to translate
Occam's razor into a mathematical language for model selection.  The
prime tool for the model selection we focus on is the Bayesian
evidence. 

The Bayesian evidence provides a natural mechanism to balance the
complexity of cosmological models and then, elegantly, incorporates
Occam's razor. It applies the same type of analysis from parameter
estimation but now at the level of models rather than parameters \citep{Trotta07}.
Let us consider several models $M$, each of them with probability $P(M)$. 
The Bayesian evidence appears again, but now 
in another form of the Bayes theorem:
\begin{equation}\label{eq:bay_model}
  P(M |\bmath{D})= \frac{P(\bmath{D}| M)P(M)}{P(\bmath{D})}.
\end{equation}
The left-hand side denotes the probability of the model given the
data, which is exactly what we are looking for model selection.  

It was previously mentioned that the Bayesian evidence is simply the
normalisation over its posterior expressed by:
\begin{equation}
  \mathcal{Z}= \int \mathcal{L}(\textbf{$\Theta$})\pi(\textbf{$\Theta$})d^N\textbf{$\Theta$},
\end{equation}
where $N$ is the dimensionality of the parameter space. 
When two models are compared, $M_0$ and $M_1$, the quantity to bear in mind is
the ratio of the posterior probabilities given by
\begin{equation}
  \frac{P(M_0 | \bmath{D})}{P(M_1 | \bmath{D})}= \,\, \frac{ \mathcal{Z}_0}{ \mathcal{Z}_1}\frac{P(M_0)}{P(M_1)} = \,\,
  B_{01}\frac{P(M_0)}{P(M_1)},
\end{equation}
where $P(M_0)/P(M_1)$ is the prior probability ratio for two models,
usually set to unity. The evidence ratio, often termed as the
{\it Bayes factor} $B_{01}$, quantifies how well model $0$ may fit
data when it is compared to model 1.  \citet{Jeffreys61} provided a
useful guide on which we are able to make qualitative conclusions
based on this difference (see Table \ref{tab:Jeffrey}).

\begin{table}
\caption{Jeffreys scale for evaluating the strength of evidence when two models are compared.}
\begin{tabular}{cccc} 
\toprule
\cline{1-4}\noalign{\smallskip}
\vspace{0.2cm}
$|\ln B_{01}|$ & Odds & Probability & \,\,\, Strength of $\mathcal{Z}$\\

\hline
\vspace{0.2cm}
$<$ 1.0 &  $< $ 3 : 1  		& $<$ 0.750   & Inconclusive \\
\vspace{0.2cm}
1.0       &   $\sim$ 3 : 1      		   & 0.750           & Significant \\
\vspace{0.2cm}
2.5      &    $\sim$ 12 : 1   		    & 0.923           & Strong \\
\vspace{0.2cm}
$>$ 5.0      &       $>$ 150 : 1  	   & $>$ 0.993           & Decisive\\
\hline
\hline
\end{tabular} 
\label{tab:Jeffrey}
\end{table}

Until recently, numerical methods such as thermodynamic integration
\citep{Beltran05, Bridges06} required around $10^7$ likelihood
evaluations to obtain accurate estimates of the Bayesian evidence,
making the procedure hardly tractable. The {\it nested sampling}
algorithm, invented by \citet{Skilling04}, has been successfully
implemented for cosmological applications \citep{Pia06,Shaw07,Feroz07}
requiring about a hundred times fewer posterior evaluations than
thermodynamic integration to achieve the same accuracy in the evidence
estimate. A substantially improved and fully-parallelized algorithm
called {\sc MultiNest} \citep{Feroz08} increases the sampling
efficiency for calculating the evidence and obtaining posterior
samples, even from distributions with multiple modes and/or pronounced
degeneracies between parameters. In our case, to carry out the
exploration of the cosmological parameter space we use a modified
version of both the {\sc MultiNest} \citep{Feroz08} and {\sc CosmoMC}
\citep{Lewis02} packages.

 \section{Datasets}
 
 Measurements of the CMB
anisotropies and large-scale structure play an important role in both
fitting parameters and comparison of models in
cosmology.
To constrain the space-parameter in each model, we first use the latest
7-year data release from WMAP (henceforth WMAP7;
\citealt{Larson10}), which includes a good measurement up to the
third acoustic peak in the temperature CMB spectrum. This comprises
our dataset 1.

WMAP7 data by itself cannot, however, place strong constraints on all
the parameters because of the existence of parameter degeneracies,
such as the $\tau- A$ degeneracy and the well-known geometrical
degeneracy, involving $\Omega_{\rm m}$, $\Omega_{\Lambda}$ and
$\Omega_k$.  Nevertheless, when WMAP7 is combined with other
cosmological observations, they together increase the constraining
power and considerably weaken the degeneracies. In addition to WMAP7,
we therefore include recent results from CMB experiments that are able
to reach higher resolution on small patches of the sky, such as the
Arcminute Cosmology Bolometer Array (ACBAR; \citealt{ACBAR}), Cosmic
Background Imager (CBI; \citealt{CBI}), Ballon Observations of
Millimetric Extra-galactic Radiation and Geophysics (BOOMERang;
\citealt{BOOM}). We also include observations of the matter power
spectrum from the luminous red galaxy (LRG) subset DR7 of the Sloan
Digital Sky Survey (SDSS; \citealt{SDSS}).  In addition to CMB and
galaxy surveys we include the Hubble Space Telescope (HST;
\citealt{HST}) key project for the Hubble parameter $H_0$. Together these
observations make up dataset 2.
\\

Throughout, we consider purely Gaussian adiabatic scalar perturbations
and neglect tensor contributions.  We assume a universe with a
cosmological constant where the background cosmology is specified by
the following five parameters: the physical baryonic matter density
$\Omega_{\rm b} h^2$, the physical dark matter density $\Omega_{\rm
 DM} h^2$, the ratio of the sound horizon to angular diameter
distance $\theta$, the optical depth to reionisation $\tau$ and the
curvature density $\Omega_k$.

The parameters defining each model are listed in
Table~\ref{tab:posteriors}, together with the ranges of the flat prior
imposed on them in our Bayesian analysis.

 \section{Parameter Estimation}
 \label{sec:params}

\begin{table}
  \caption{Parameters and prior ranges used in our analysis.}
 \begin{tabular}{cccc}
\toprule
\cline{1-4}\noalign{\smallskip}
\qquad Model  \qquad & Parameter	 \qquad	&\qquad Prior range \\
\hline
{All$^{\phantom{\int^\int}}$} 
& $\Omega_{\rm b} h^2$  	 		& 0.01 & 0.03 \\
&$\Omega_{\rm DM} h^2$   		  &  0.01 & 0.3\\
&$\theta$               				&  1.0 & 1.1 \\		     		
&$\tau$               				& 0.01 & 0.3\\
\hline
{Power-law$^{\phantom{\int^\int}}$} 
&$\log[10^{10}A_{\rm s}]$        		  & 2.5 & 4.0 \\
PL 1
&$n_{\rm s}$                             		 & 0.5 & 1.5\\ \cline{2-4}
PL 2
&$\Omega_k$                  			 & -0.2 & 0.2  \\ \cline{2-4}
PL 3 &$n_{\rm run}$                  	 & -0.2 & 0.2  \\
\hline
{LD $^{\phantom{\int^\int}}$} 
&$\Omega_k$          				& -0.2 & -0.0001\\
&$b_0 $                       			& 1.0 & 4.0\\
&$b_4 [10^{-9}]  $      			 & -30.0 & -0.1\\
\bottomrule
\end{tabular}
\label{tab:posteriors}
 \end{table}

  \subsection{Power-law parameterisations}
 
  \begin{figure} 
   \includegraphics[trim = 2mm 109mm 5mm 110mm, clip, width=9cm, height=2.5cm]{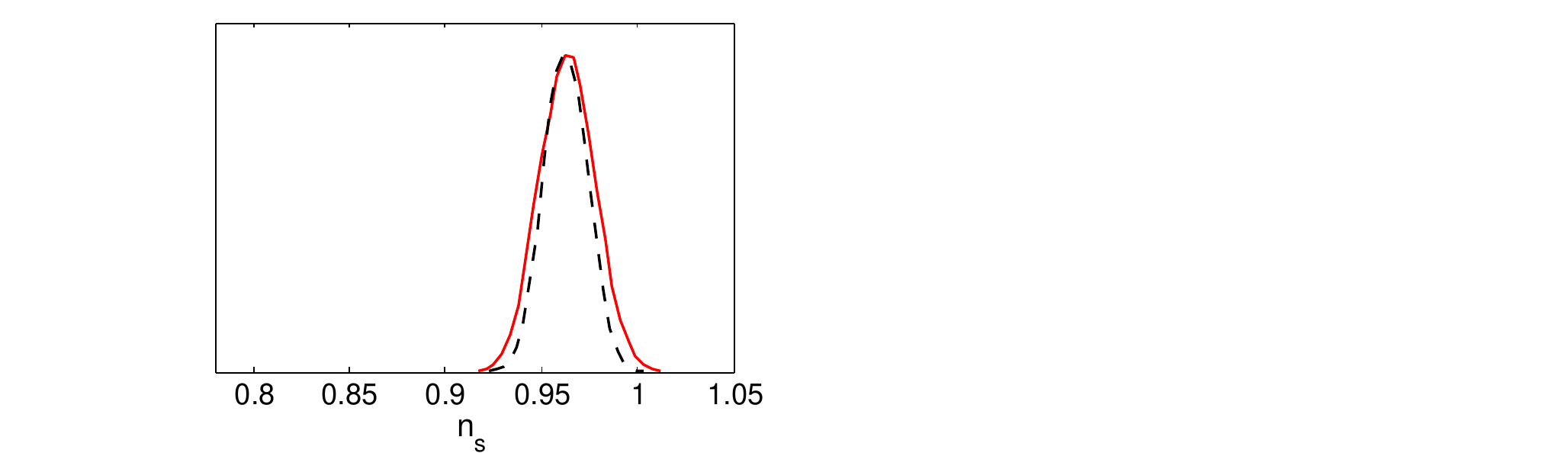}
    \includegraphics[trim = 2mm 109mm 5mm 110mm, clip, width=9cm, height=2.5cm]{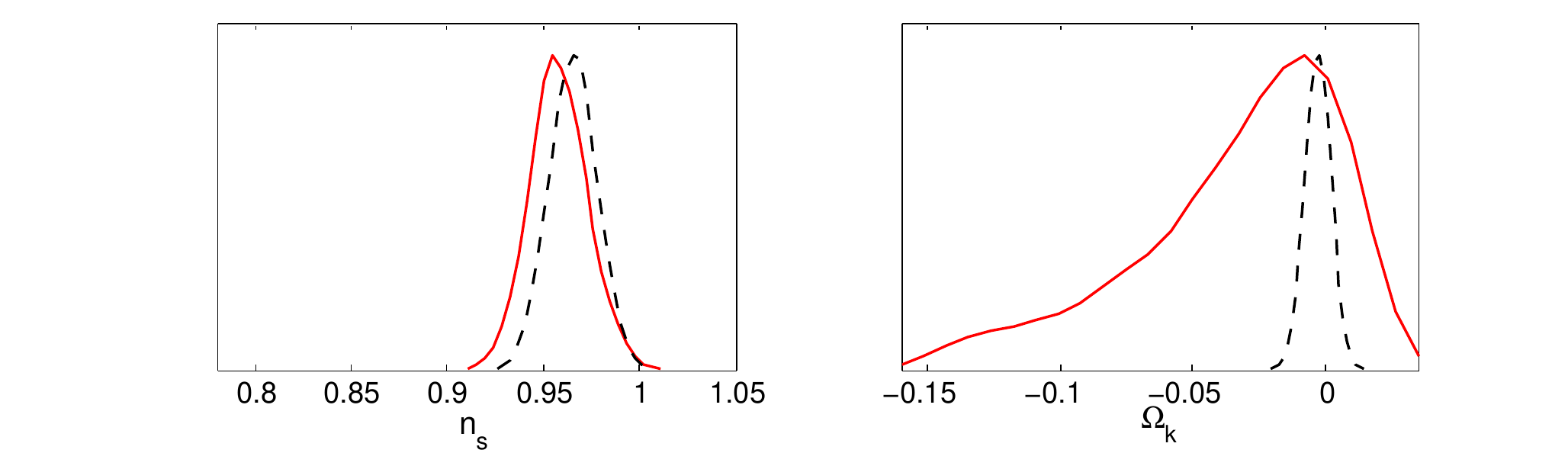}
     \includegraphics[trim = 2mm 109mm 5mm 110mm, clip, width=9cm, height=2.5cm]{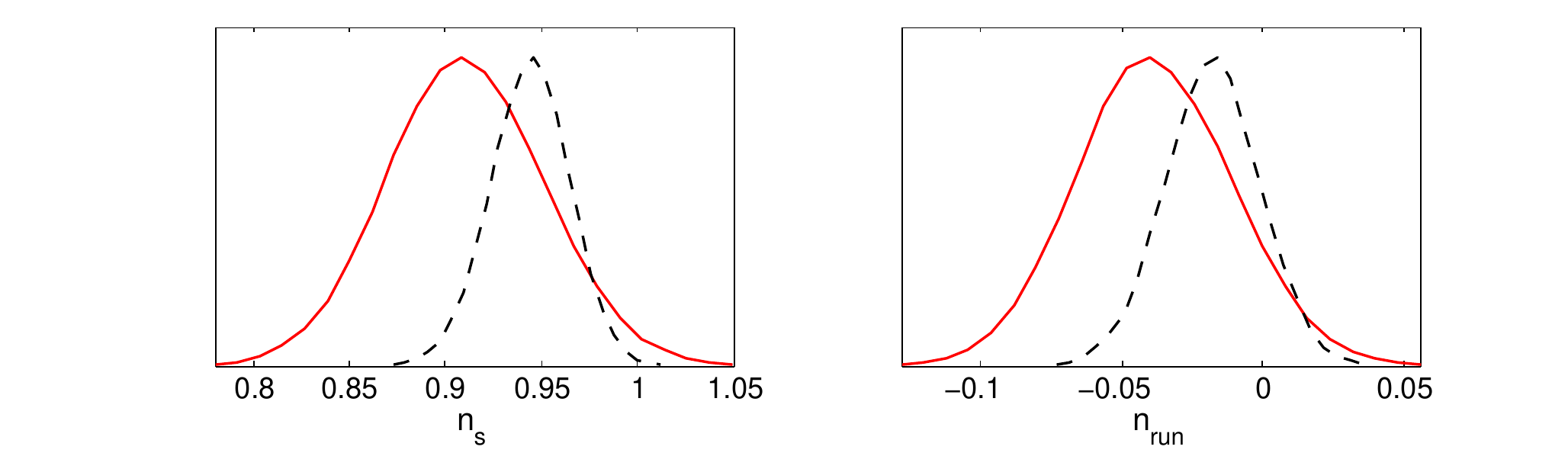}
    \caption{1-D marginal posterior distributions for different power
      spectrum parameter models: PL 1 ($n_{\rm s}$; top row), PL 2 
($n_{\rm s} + \Omega_k$; middle row) and PL 3 ($n_{\rm s}+n_{\rm run}$; bottom row), using 
dataset 1 (solid line) and dataset 2
      (dashed line).}
    \label{fig:index}
   \end{figure}

In Fig \ref{fig:index} we plot the 1-D marginal posteriors for the
power spectrum parameters in each of the power-law models, using
both dataset 1 and dataset 2. 
We now on refer to the mean of the posterior distribution of each parameter
together with its 68 \% confidence interval.
For model 1 ($n_{\rm s}$), we see that
scale-invariant spectrum $n_{\rm s}=1$ is ruled out, as expected,  with high 
confidence level using either dataset. Indeed, the constraints
on $n_{\rm s}$ for a flat universe (PL 1) and a curved universe (PL 2) 
using dataset 2 are similar: $n_{\rm s}=0.963\pm
0.011$ and $n_{\rm s}=0.965 \pm 0.012$ respectively.  The inclusion of the
running parameter (PL 3) displaces and broadens the posterior probability,
such that $n_{\rm s}=0.944 \pm 0.020 $.

When the curvature is considered as a free parameter, dataset 1 
selects a closed universe 
$\Omega_k=-0.033 ^{+0.030}_{-0.041}$ with a Hubble constant
$H_0=61.0^{+12.2}_{-12.5} $ km/s/Mpc. Similarly, for a flat Universe the constraints for the running parameter
 $n_{\rm run} =-0.038 \pm  0.027$ and $H_0=66.6\pm3.8$ km/s/Mpc
 are in good agreement with \citet{Larson10}.
 The inclusion of measurements on
different scales (dataset 2) weakens the geometric degeneracy and
tightens significantly the constraints, yielding a mean posterior
value for the curvature $\Omega_k=-0.0026\pm 0.0049$ and Hubble
parameter $H_0=70.4\pm 1.7$ km/s/Mpc. On the other hand, we observe the 
best-fit for the running parameter is moved closer to its zero value with constraints 
$n_{\rm run} =-0.018 \pm  0.016$ 

\begin{figure}
  \includegraphics[trim = 1mm 0mm -10mm 1mm, clip, width=9cm, height=8.5cm]{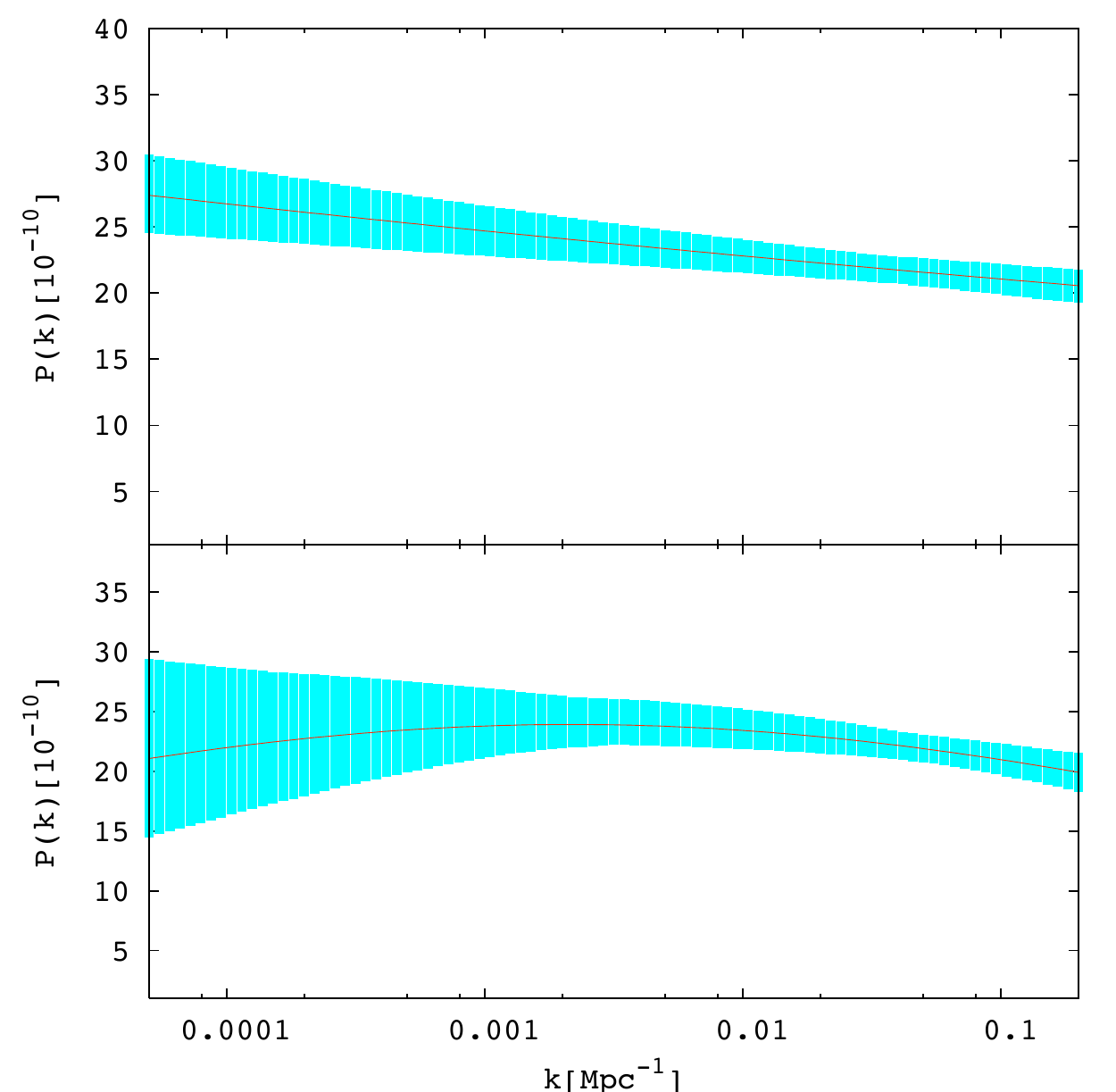}
	\caption{Primordial power spectrum reconstructed using mean values of the posterior distributions found from
	dataset 2, for PL model 2 ($n_{\rm s}+\Omega_k$; top panel)
	and PL model 3 ($n_{\rm s}+n_{ \rm  run}$; bottom panel), with 
	$1\sigma$ error bands.}
	\label{fig:pri_power}
\end{figure}

In Fig \ref{fig:pri_power}, we plot the reconstructed  shape  for the power
spectrum for the simple spectral index  in  a 
 curved universe and the running parameterisation in flat universe,
using mean values of the posterior distributions found from dataset 2.

\subsection{LD model}

\begin{figure}
   \includegraphics[trim = 10mm 50mm 10mm 50mm, clip, width=9cm, height=8cm]{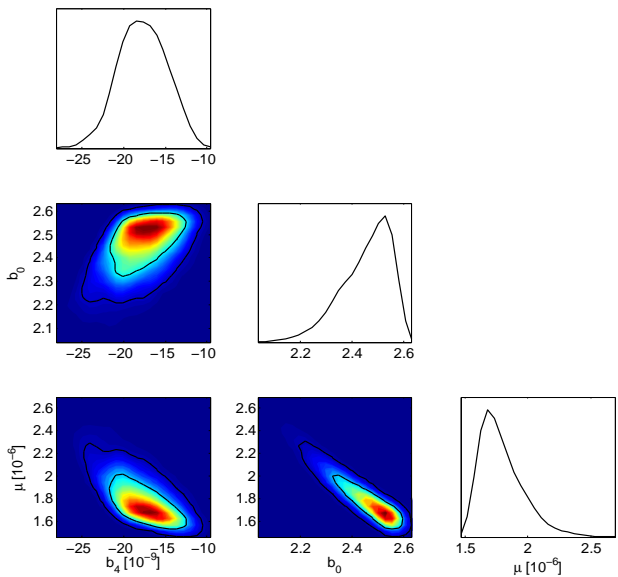}
    \caption{Marginalised 1D and 2D probability distributions for the
      power spectrum parameters $b_4$, $b_0$ and $\mu$ in the LD
      model, using dataset 2.  2D constraints are plotted with
      $1\sigma$ and $2\sigma$ confidence contours.}
    \label{fig:inf}
  \end{figure}

We choose to describe the LD spectrum in terms of the parameters $b_0$
and $b_4$, letting $\mu$ be a derived parameter such that the
conformal time constraint is fulfilled.  The resulting constraints
from dataset 2 on the parameters describing the LD power spectrum are
shown in Fig \ref{fig:inf}.
%
In particular we obtain the constraint $b_0=2.45^{+0.90}_{-0.11}$,
which is intimately linked to the number of $e$-folds $N$ during the
inflationary epoch through (\ref{eq:Nb0}).  
We also find the constraint $b_4\,
[10^{-9}]=-17.74^{+2.92}_{-2.78}$ and we note the effect occurring  
through (\ref{eq:ini_cur}): a greater initial curvature (increased
$|b_4|$) would drive the universe today closer to be flat (decrease
$|\Omega_k - 1|$).  Finally, the constraint on the scalar field mass
is $\mu \, [10^{-6}]=1.79^{+0.17}_{-0.16}$.

\begin{figure}
   \includegraphics[trim = 10mm 60mm 3mm 60mm, clip, width=9cm, height=6cm]{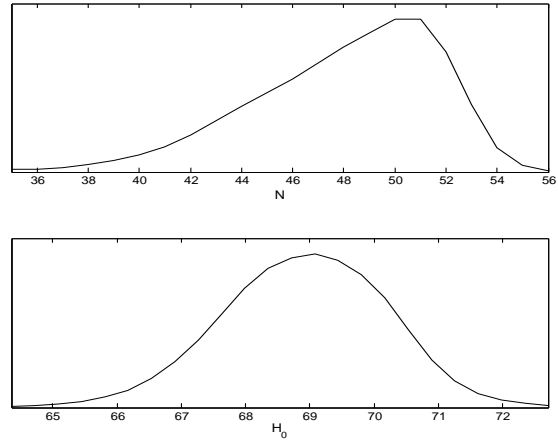}
  \caption{Marginalised 1D probability distributions for the number of
    e-foldings produced during inflation (top) and Hubble constant (bottom), obtained from the LD
    model using dataset 2.}
 \label{fig:folds}
\end{figure}

The LD model requires a closed universe, then using dataset 2 a very small
value for the curvature parameter is obtained
$\Omega_k=-0.43^{+0.27}_{-0.28}\times 10^{-2}$, together with a Hubble constant
of $H_0=68.9\pm1.3$ km/s/Mpc, both of which are compatible with all
existing observations.  We note, however, that many authors have
argued on the difficulty to construct realistic closed-universe models
with the required number of $e$-folds and also that a fine tuning might arise from the initial
conditions to obtain an homogeneous universe after inflation
\citep{Ellis02, Linde03, Uzan03}.  In this sense, the LD model
provides the correct order of magnitude in the prediction of the
number of $e$-folds given by $N=48.1^{+3.3}_{-4.2}$ (see Fig
\ref{fig:folds}).

\begin{figure}
  \centerline{ \epsfxsize=240pt \epsfbox{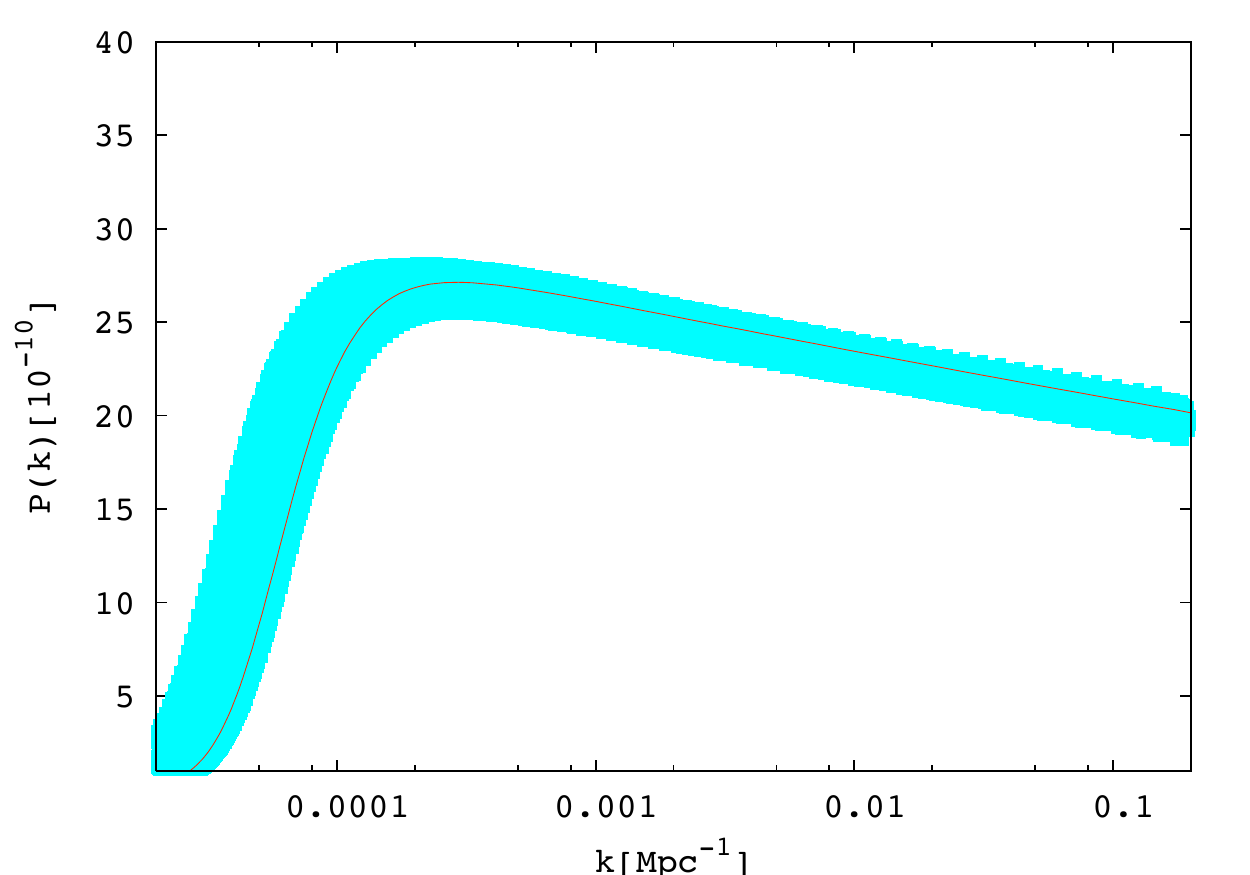} }
	\caption{Primordial power spectrum reconstructed from the LD model 
using mean values of the posterior  distributions obtained from dataset 2, 
	  with $1\sigma$ error bands.}
\label{fig:Power}	
\end{figure}

Finally, we plot in Fig.~\ref{fig:Power} the reconstructed primordial
power spectrum for the LD model together with $1\sigma$
error bands. It is worth pointing out that the LD 
 spectrum automatically contains a clear cut-off at low
$k$-values and an additional fall-off at high wavenumber. A low-$k$
truncation of the primordial spectrum of this type was proposed
phenomenologically by \citet{Efstathiou03}.  This kind of behaviour
could be responsible for the small CMB power currently observed at low
$l$ multipoles.

\section{Model selection}

We now investigate which model provides the best description of
the data by performing model selection based on the value of the
Bayesian evidence. Before performing this process on the real
datasets, however, we first test our approach by applying 
it to an idealised dataset.

\subsection{Application to simulated data}
 
  \begin{figure}
   \centerline{ \epsfxsize=240pt \epsfbox{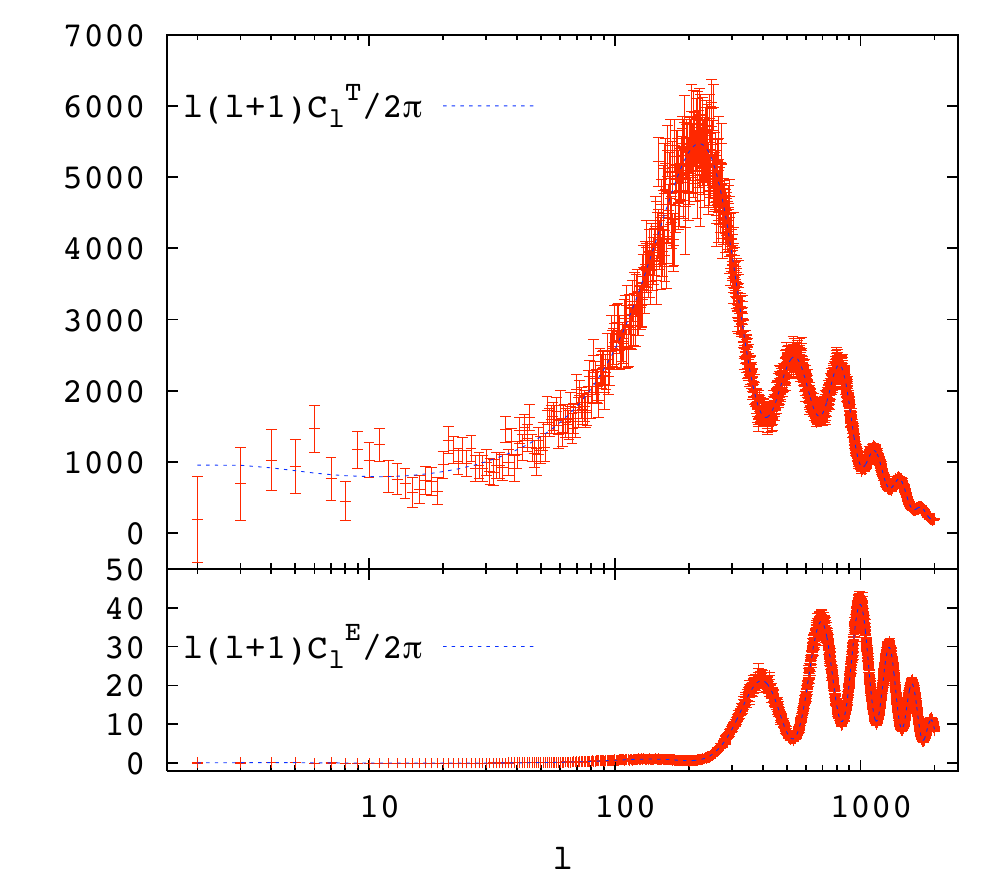} }
    \caption{Top (bottom) panel shows a simulated  
    CMB temperature (polarisation) power spectrum, convolved with chi-square noise and limited data up to $l$ of 2000.
     The model from which it was produced is drawn by the blue line.}
	\label{fig:noise}
   \end{figure}

To test our model selection (and parameter estimation) method, we
implement the LD spectrum in a modified version of the {\sc CAMB} package
\citep{Lewis00} to simulate the CMB power spectrum predicted by the LD
model using standard values for the cosmological parameters, i.e.  those obtained  from  WMAP7.
  The predicted temperature and E-mode polarisation CMB spectra are plotted
as the dashed line in Fig.~\ref{fig:noise}. We simulate an
idealised process where  only cosmic variance noise was added to the spectra such that
each $C_l$ value becomes a random variable drawn from a 
$\chi^2_{2l+1}$ distribution with variance
  \begin{equation}
 (\Delta C_l)^2 = \frac{2}{2l+1}C_l^2.
\end{equation}

\begin{figure}
\includegraphics[trim =30mm 94mm 25mm 95mm, clip, width=9cm, height=5.9cm]{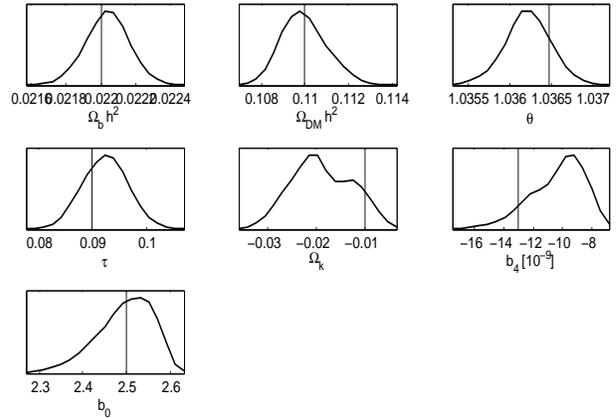}
\caption{1-D marginalised parameter constraints for a simulated model. The set of parameters used to 
construct the mock data is represented by the vertical lines. The $y$-axis is relative probability.}
\label{fig:pos_noise}
\end{figure}

The 1D marginalised parameter constraints for the LD model resulting
from the analysis of this simulated dataset are shown in Figure
\ref{fig:pos_noise}. 
We observe 
that our approach yields constraints that appear consistent, within statistical error, 
with the input values used in the simulation, which are indicated by the vertical lines
in each plot. Moreover, these
results give us an indication of how accurately the parameters of
the LD model can be constrained with an optimal CMB dataset. In
particular, we note that the behaviour of the posterior for $\Omega_k$
is reflected in that obtained for $b_4$, that is because of the link between
them through the see-saw mechanism.

The evidence values of the models considered for the primordial 
spectrum are given in Table \ref{tab:noise}.  Based on Jeffreys' criterion we observe
significant difference in the log-evidences for each model.  Thus,
there is a clear distinction between models, with the data clearly
indicating a preference, as expected, for the LD model used to
generate the simulated data. Given the success of model selection on
simulated data, we now turn to real data with some confidence.

\begin{table}
\caption{Difference of log-evidences for the different
  parameterisations relative to simple power-law flat model for the
  simulated CMB dataset.}
\begin{tabular}{lc} 
\toprule
\cline{1-2}\noalign{\smallskip}
\vspace{0.2cm}
Model &  \qquad   $\Delta \ln \mathcal{Z}$\\

\hline 
PL  1 ($n_{\rm s}$)					 \qquad 	& \qquad    0.0 \, $\pm$ \, 0.4\\
PL  2 ($n_{\rm s}$ + $\Omega_k$)		 \qquad 	&  \qquad   -1.0 \, $\pm$ \,  0.4\\
PL  3 ($n_{\rm s}$ + $n_{\rm run}$)	 	\qquad 	&  \qquad  -3.1 \, $\pm$ \,  0.4\\
LD model   						\qquad 	&   \qquad  +3.7 \, $\pm$ \,  0.4 \\
\hline	
\hline
\end{tabular} 
\label{tab:noise}
\end{table}

\subsection{Application to real data}

Before performing the Bayesian selection between our different models
for the primordial power spectrum, we will momentarily use a
frequentist method and compute the goodness-of-fit of each of the models
using real data. Using the best-fit parameters obtained for each model
with dataset 1, we may calculate $\chi^2_{\rm min}$, defined as $- 2\ln
\mathcal{L}_{\rm max}$, at the maximum point in each case. These
values are given in Table~\ref{tab:chi} from which we see that all
three models fit the data almost equally as well.
 \begin{table}
 \caption{Log-likelihood values at the best-fit point for
three different models of the primordial power spectrum using dataset 1.}
\begin{tabular}{lcc} 
\toprule
\cline{1-3}\noalign{\smallskip}
Model & $N_{\rm params}$ & -$\ln \mathcal{L}_{max}$  \\
\hline
PL  2 ($n_{\rm s}$ + $\Omega_k$)			& 7 		& 7475.3 	\\
PL  3 ($n_{\rm s}$ + $n_{\rm run}$)	 		& 7 		& 7473.5 	\\
LD model							  	& 7		& 7473.0	\\
\hline
\hline
\end{tabular} 
\label{tab:chi}
\end{table} 

Turning now to Bayesian model selection, the difference in the
log-evidences for each of the models are
given in Table~\ref{tab:evidence}, as derived from dataset 1 and
dataset 2 respectively.
\begin{table}
 \caption{Difference of log-evidences for each of our models for the
   primordial power spectrum, relative to the simple power-law model.}
 \begin{tabular}{lcrr} 
 \toprule
 \cline{1-4}\noalign{\smallskip}
 \vspace{0.2cm}
Model & $N_{\rm params}$ &  Dataset 1 & Dataset 2 \\
\hline
PL  1 ($n_{\rm s}$)					& 6 & 	0.0 $\pm$ 0.3 			& 0.0 $\pm$ 0.3 \\
PL  2 ($n_{\rm s}$ + $\Omega_k$) 		& 7 & 	$-0.5$ $\pm$ 0.3 		& $-2.6$ $\pm$ 0.3 \\
PL  3 ($n_{\rm s}$ + $n_{\rm run}$)	 	 & 7 &	$-0.8$ $\pm$ 0.3		 & $-1.7$ $\pm$ 0.3 \\
LD model  						& 7 &  	$+1.2$ $\pm$ 0.3 		 & $-0.9$ $\pm$ 0.3 \\
\hline
\end{tabular} 
\label{tab:evidence}
\end{table}
The log-evidence difference should be interpreted in the context of
the Jeffreys' scale given in Table~\ref{tab:Jeffrey}.  
In the analysis we have  considered a wide conservative prior for the parameters in each model. 
The parameter  estimation results in Section \ref{sec:params}  showed 
that the constraints  obtained lie well within our chosen prior and as such any increase in the prior would
not include very much more posterior mass within the evidence integral. Thus we would not expect 
reasonable modifications of the prior range to alter the results makedly.

For dataset 1, the similarity in log-evidences and their statistical
uncertainties make it difficult to obtain definitive conclusions.
When the evidence values are ranked, however, we observe a slight
preference in favour of the LD spectrum over the three power-law
parameterisations, whereas model 3 ($n_{\rm s} + n_{\rm run}$) is
significantly the most disfavoured.  The inclusion of additional
information on different scales with dataset 2 significantly reduces
the evidence for the three 7-parameter models.  We notice that a curved universe with a simple power-law, PL 2
($n_{\rm s}+\Omega_k$), is strongly disfavoured relative to a flat universe with a power-law spectrum
 ($n_{\rm s}$).  When we perform a comparison of models which contain
 the same number of parameters, we observe the LD model is 1.7
units of log-evidence above the PL model 2, which is a significant
difference, but only 0.8 log-units above PL model 3. The $\Lambda$CDM model with a
 power-law spectrum  has the largest evidence in this case. We observe from Table \ref{tab:evidence}, 
that from all of the models considered with equal numbers of parameters (with either dataset) that the LD model is currently preferred.

\section{Discussion and Conclusions}
 
 The novel closed universe model proposed by Lasenby $\&$ Doran
 gives rise to a boundary condition on the
 total available conformal time:
   \begin{equation*}
\eta_{M}+\eta_{I}= \frac{\pi}{2}. 
\end{equation*}
Here we have split the universe evolution in two main epochs that
contribute to the amount of conformal time: matter-dominated epoch and
inflationary period.  The matter epoch can be described in terms of
matter-energy content of the universe: baryons ($\Omega_{\rm b},$), photons
($\Omega_{\rm r}$), Cold dark matter ($\Omega_{\rm DM}$) and dark energy
($\Omega_{\Lambda}$), as well as on its expansion history: Hubble time
($H_0$). The inflationary period is encoded in three
parameters that have emerged in the construction of the appropriate
initial conditions.  They are related to the initial curvature
($b_4$), the initial expansion size of the universe ($b_0$) and the
amplitude of perturbations ($\mu$).  Such parameters are not predicted
a priori, rather we have fit them using current observational data. 
Notice that we have the freedom
to pick two of them and compute the third parameter through the constraint imposed 
on the conformal time. In this paper we have focused on $b_0$ and $b_4$ to
describe the LD spectrum.  
The comparison of the best-fit primordial spectra for our different models is shown on
Fig \ref{fig:Pri_final}.

We test the LD spectrum by building a simple toy model from a
chosen CMB spectrum with simulated cosmic variance noise.  A
full Bayesian analysis was performed for this toy model and various
parameterisations of the primordial power spectrum. We  have included not
only the estimation of cosmological and spectral parameters but also
the Bayesian evidence.
 For real data, we compared the $\chi^2$ for the three different
 models and concluded that the goodness-of-fit for the LD model is about the
 same as the well-studied power-law models.  We also pointed out that
 the constraints on the cosmological parameters  in the LD model are consistent
  within  $1\sigma $  error  bars
  with the power-law models, as shown in Figs {\ref{fig:power} and
   {\ref{fig:model}.  
   \\

\begin{figure}
\includegraphics[trim = 1mm -5mm 5mm -5mm, clip, width=8cm, height=6cm]{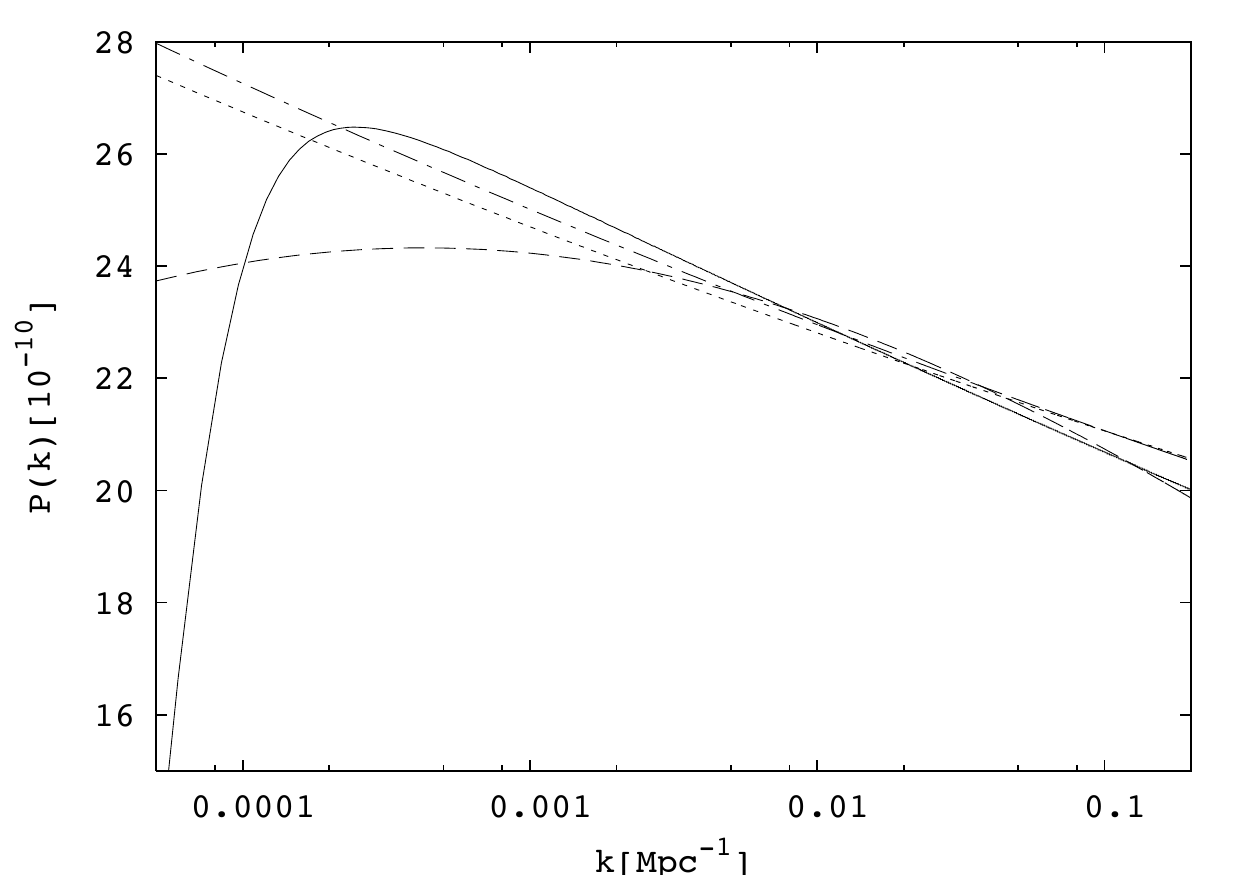}
\includegraphics[trim = 1mm -5mm 5mm -1mm, clip, width=8cm, height=6cm]{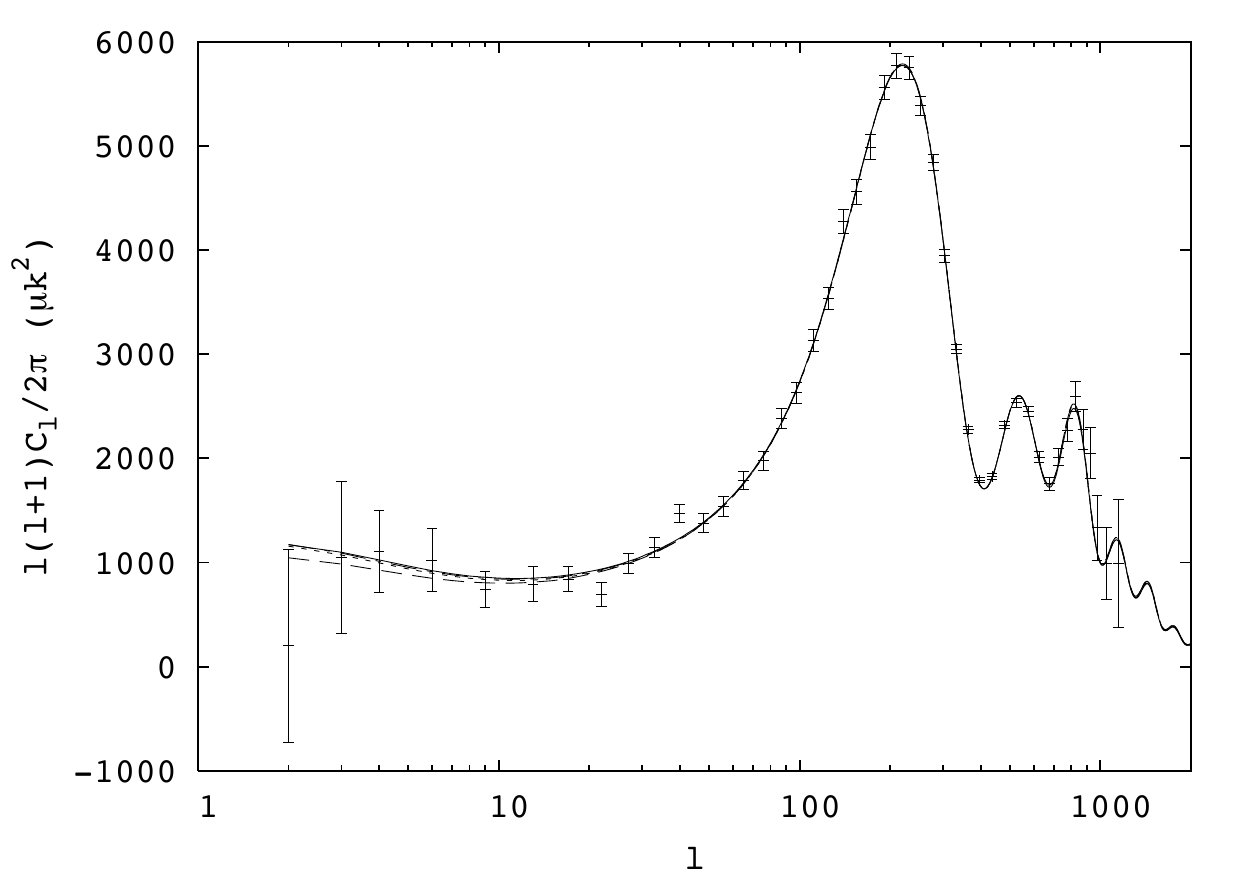}
 \caption{Primordial and anisotropy power spectrum for different
      models of the primordial spectrum: $n_{\rm s}$ (dotted line), $n_{\rm s}+\Omega_k$
      (dash-dot line), $n_{\rm s}+n_{\rm run}$ (dashed line) and LD model (solid line), using
      the best-fit parameter values for dataset 2.}
 \label{fig:Pri_final}
\end{figure}

The Lasenby $\&$ Doran model predicts a primordial scalar spectrum that at
large scales naturally incorporates a drop off without the need to
parameterise it, while, at small scales it automatically incorporates a degree of
negative running compared with the standard power law
parameterisation (see e.g. Fig \ref{fig:Pri_final} below or Fig. 17 in Lasenby $\&$ Doran, 2005, for a visual
comparison). Given that the primordial parameters $b_0$ and $b_4 /
\mu^{4/3}$ are well constrained and of order unity, no fine tuning is
needed for the construction of the initial conditions. Moreover, the LD
model has the required order of $e$-foldings to obtain realistic models
of a closed inflationary universe. Finally, even though the Bayesian
evidence from each model showed that it is difficult to make definitive
conclusions because of the presence of statistical uncertainties, an
important result to emphasise is that for the models with the same
number of parameters, the preferred model to explain current
cosmological observations in given by the LD model.
This may be seen by noting that, for models  with 7 parameters, Table  
\ref{tab:evidence} shows that the LD model is preferred significantly (by 1.7-2.0
$\log$ units of evidence) in all cases except relative to  $n_{\rm s}+n_{\rm run}$
with dataset 2, for  which it is preferred by 0.8 $\log$ units.

\section*{Acknowledgments}

This work was carried out mainly on the Cambridge High Performance
Computing Cluster Darwin.  JAV is supported by CONACYT M\'exico.

\bibliographystyle{newapa}
  \bibliography{refs}

\begin{figure*} 
\includegraphics[trim = 1mm 100mm 1mm 110mm, clip, width=18cm, height=5cm]{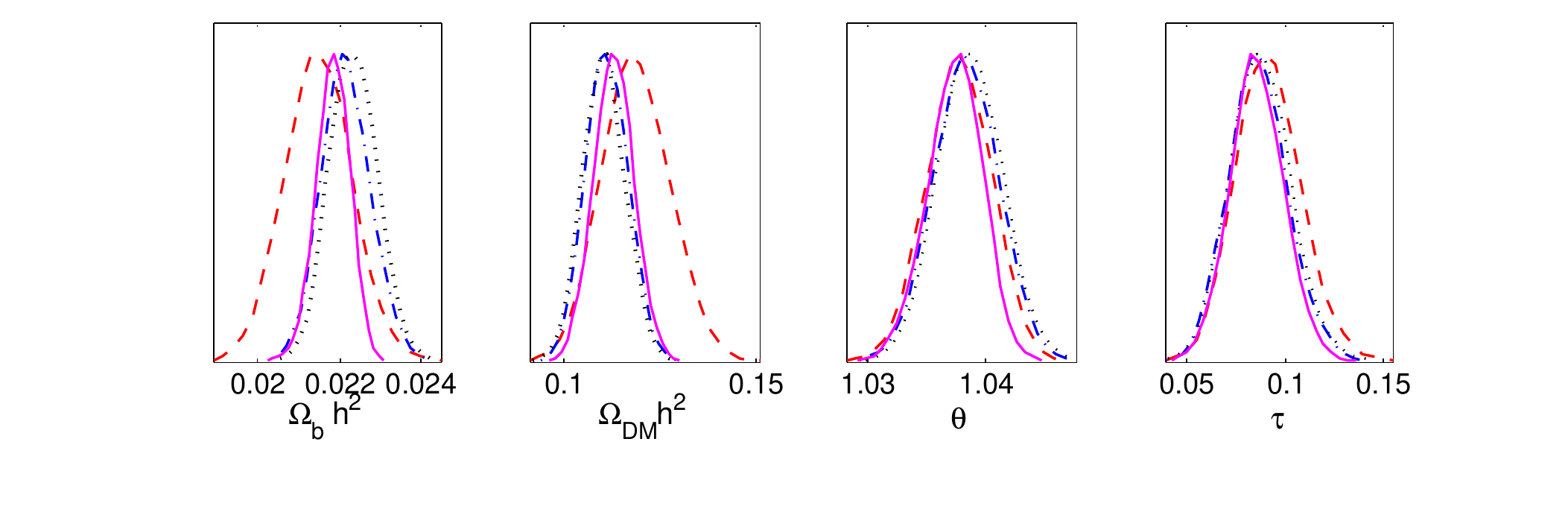}
\includegraphics[trim = 1mm 100mm 1mm 110mm, clip, width=15cm, height=4cm]{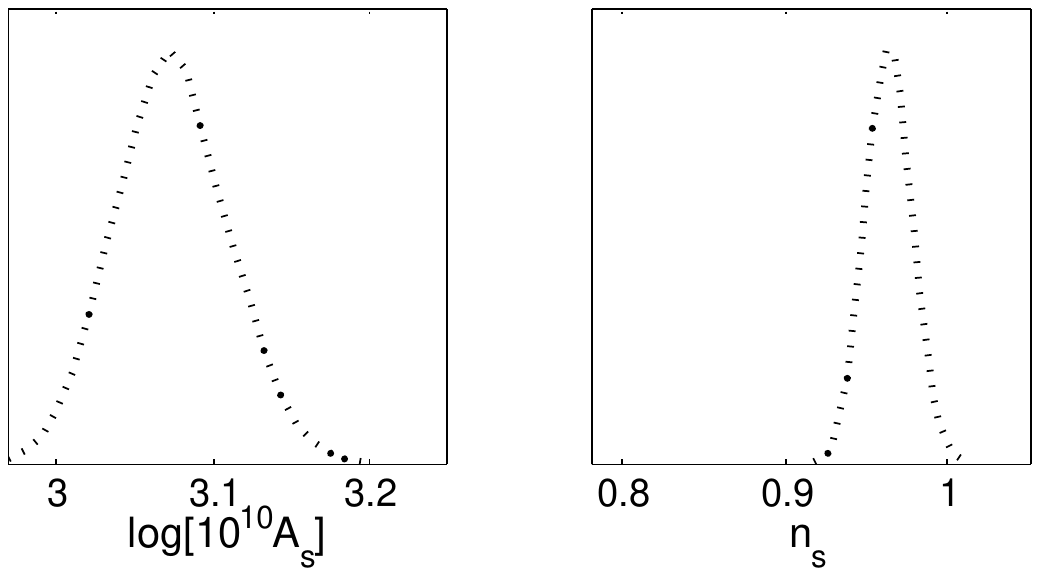}
\includegraphics[trim = 1mm 100mm 1mm 110mm, clip, width=15cm, height=4cm]{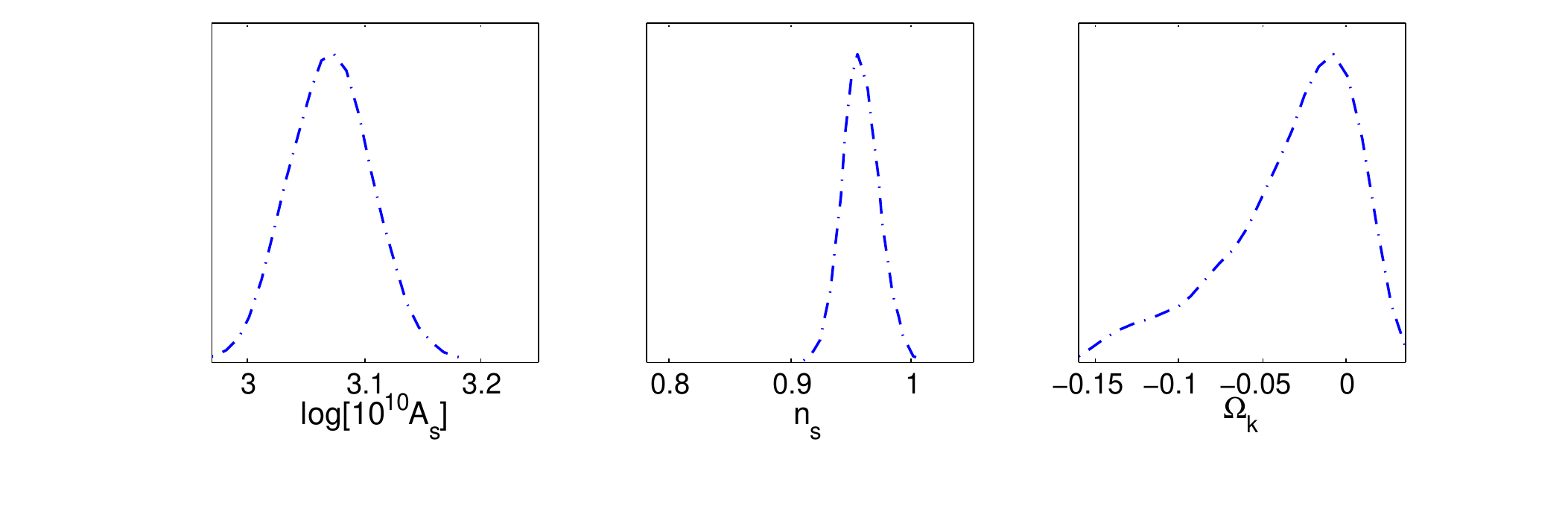}
\includegraphics[trim = 1mm 100mm 1mm 110mm, clip, width=15cm, height=4cm]{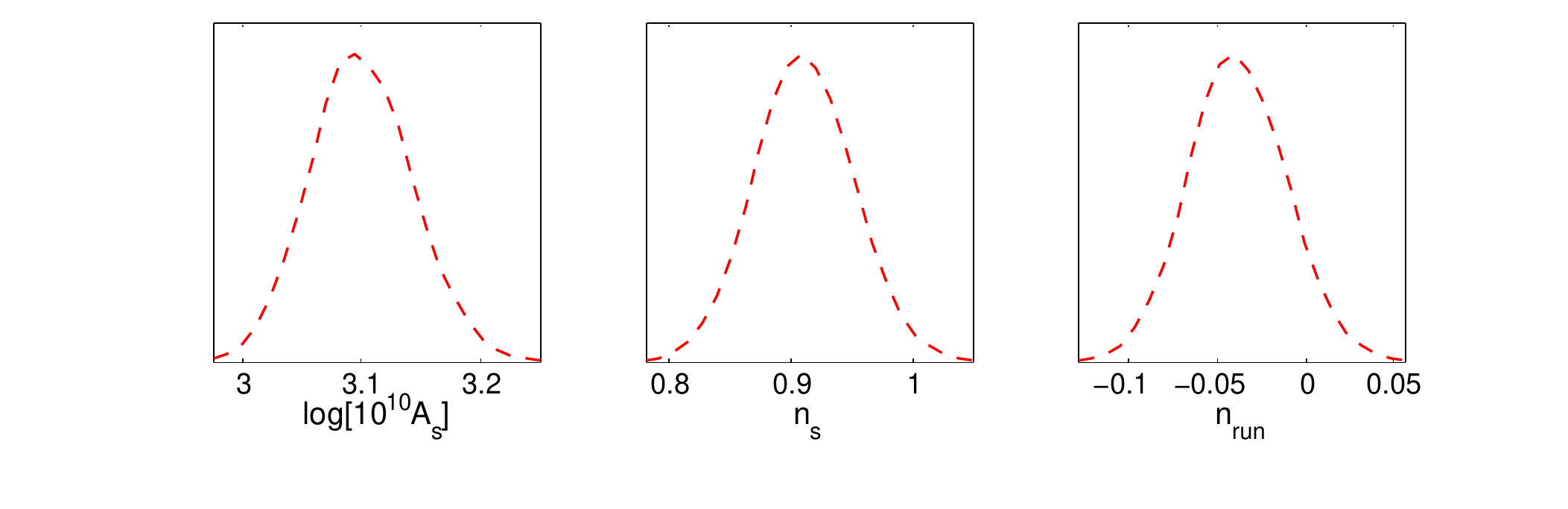}
\includegraphics[trim = 1mm 100mm 1mm 110mm, clip, width=15cm, height=4cm]{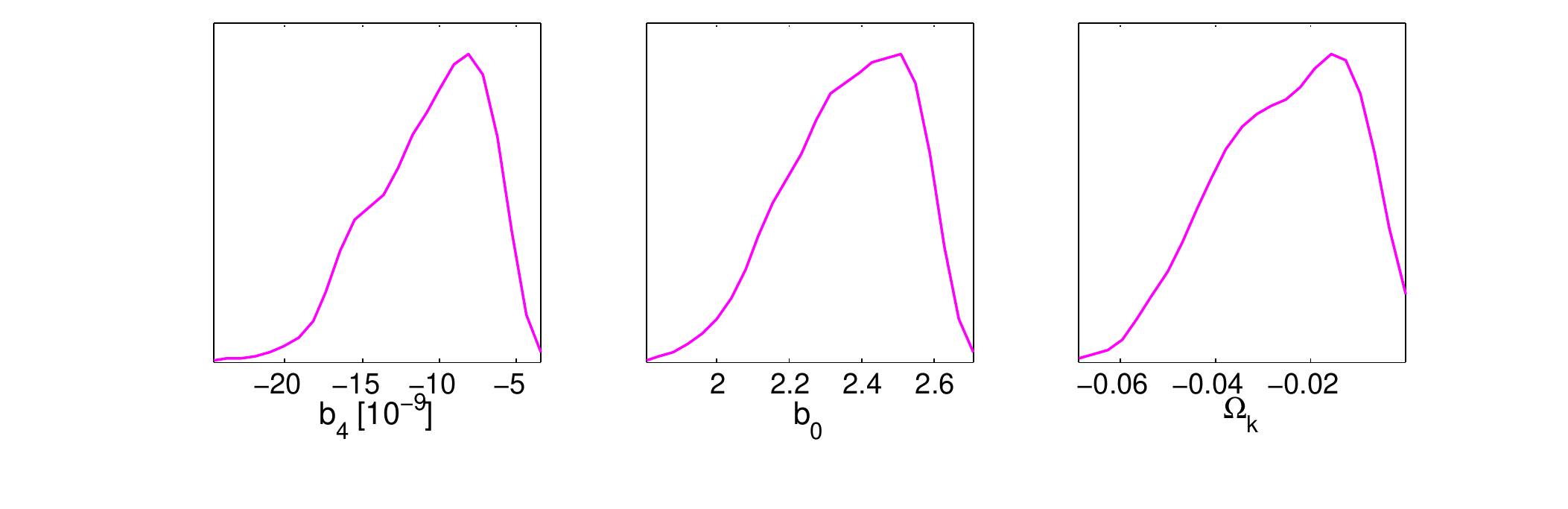}
\caption{ Marginalised parameter constraints corresponding to 
$n_{\rm s}$ (dotted line), $n_{\rm s}+\Omega_k$ (dash-dot line), $n_{\rm s}+n_{\rm run}$ (dashed line) and
    L \& D model (solid line), using dataset 1.}
\label{fig:power}
\label{lastpage}
\end{figure*}

\begin{figure*} 
\includegraphics[trim = 1mm 100mm 1mm 110mm, clip, width=18cm, height=5cm]{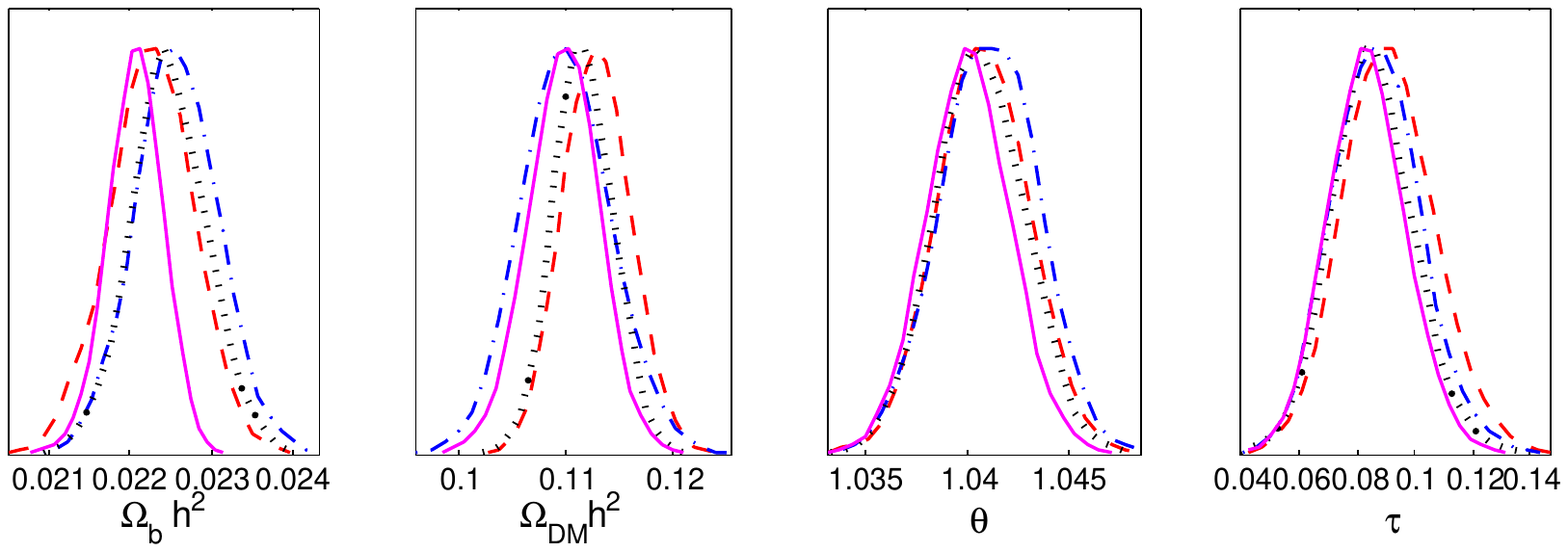}
\includegraphics[trim = 1mm 100mm 1mm 110mm, clip, width=15cm, height=4cm]{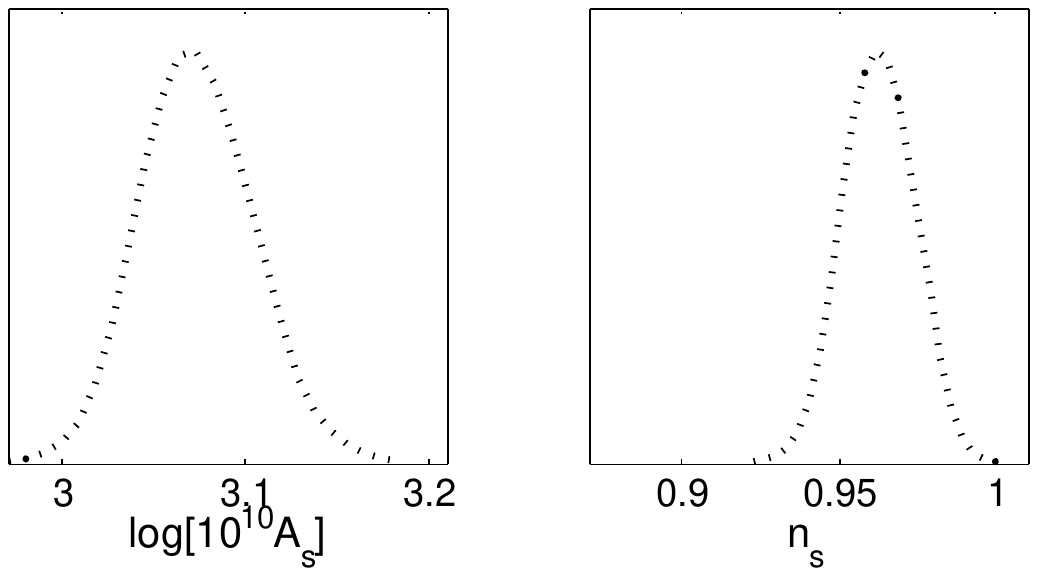}
\includegraphics[trim = 1mm 100mm 1mm 110mm, clip, width=15cm, height=4cm]{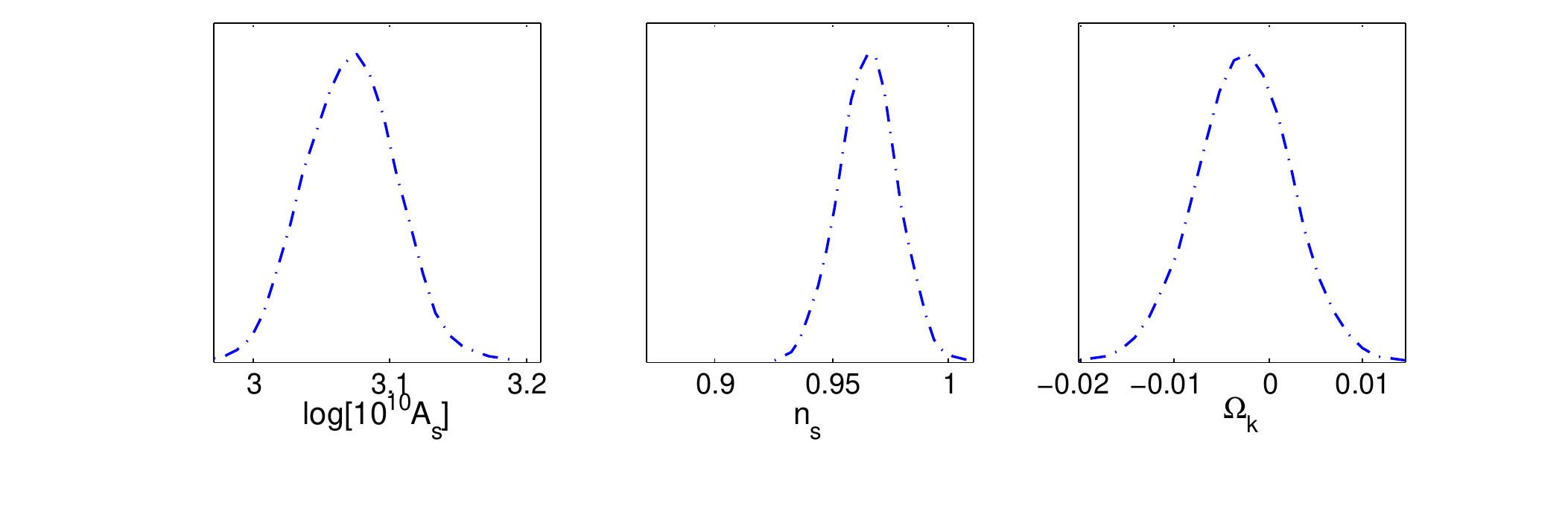}
\includegraphics[trim = 1mm 100mm 1mm 110mm, clip, width=15cm, height=4cm]{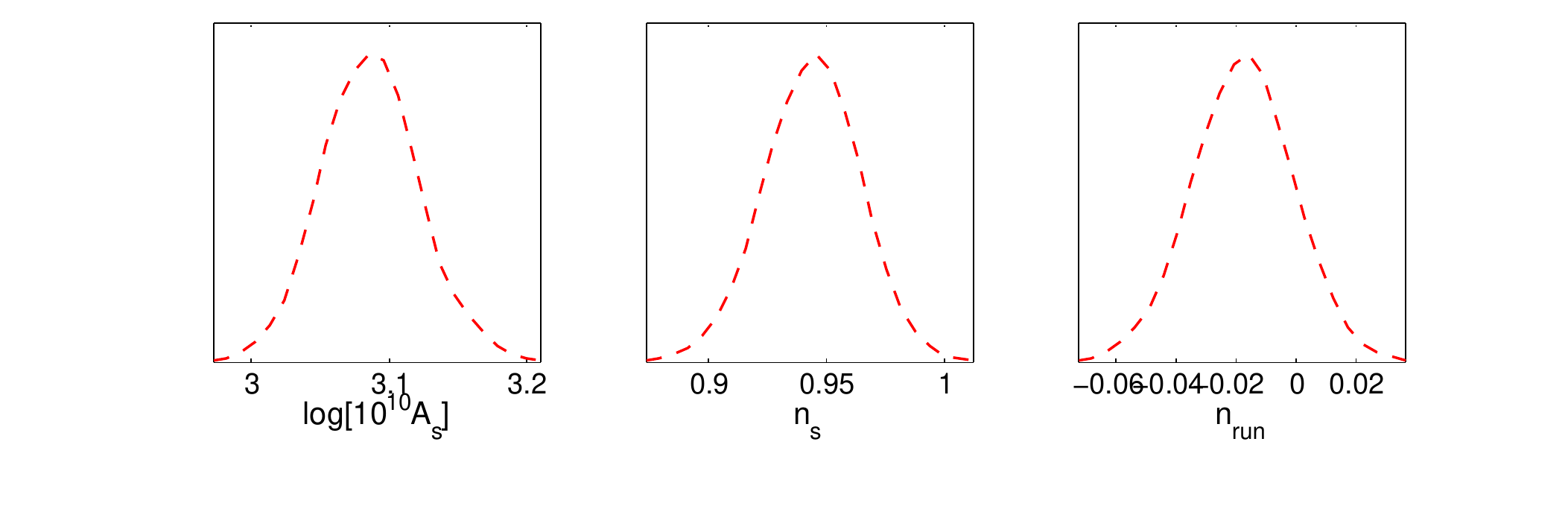}
\includegraphics[trim = 1mm 100mm 1mm 110mm, clip, width=15cm, height=4cm]{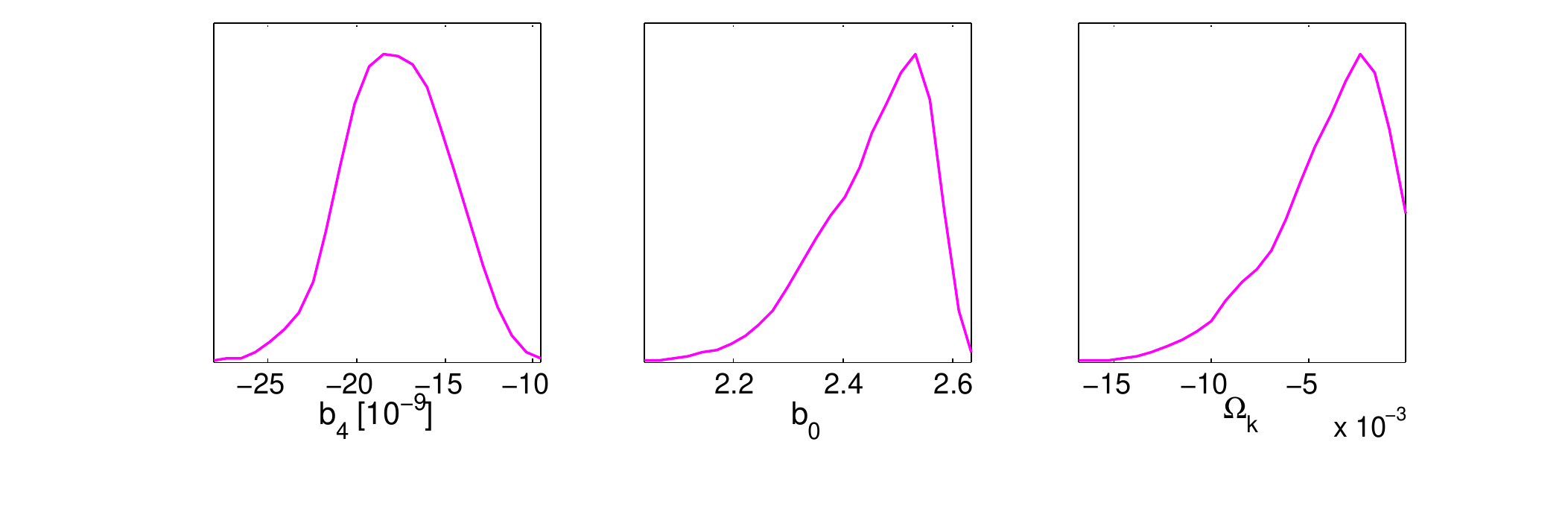}
\caption{ Marginalised parameter constraints corresponding to 
$n_{\rm s}$ (dotted line), $n_{\rm s}+\Omega_k$ (dash-dot line), $n_{\rm s}+n_{\rm run}$ (dashed line) and
    L \& D model (solid line), using dataset 2.}
\label{fig:model}
\label{lastpage}
\end{figure*}

\end{document}